\documentclass[sigconf,natbib,screen=true, nonacm]{acmart}

\usepackage[skip=0pt]{caption}
\usepackage{booktabs}
\usepackage[inline]{enumitem}
\usepackage{balance}

\usepackage{xcolor}

\usepackage{verbatim}

\usepackage{multirow}

\usepackage{numprint}
\npthousandsep{,}
\npdecimalsign{.}

\usepackage{graphicx} 
\graphicspath{{img/}} 

\usepackage{acronym}
\acrodef{IR}{information retrieval}

\usepackage{subfigure}

\usepackage{mathrsfs}

\usepackage{amsfonts}
\usepackage{amsmath}

\usepackage{hyperref}

\copyrightyear{2022}
\acmYear{2022}
\setcopyright{rightsretained}
\acmConference[]{}
\acmBooktitle{}
\acmDOI{}
\acmISBN{}

\newcommand{\header}[1]{\vspace{1mm}\noindent\textbf{#1}.}

\AtBeginDocument{%
  \providecommand\BibTeX{{%
    \normalfont B\kern-0.5em{\scshape i\kern-0.25em b}\kern-0.8em\TeX}}}

\acrodef{SOTA}{State-Of-The-Art}
\acrodef{IR}{\textbf{Information Retrieval}}
\acrodef{FITB}{\textit{Fill in the Blank}}
\acrodef{MRR}{\textit{mean reciprocal rank}}
\acrodef{GNN}{\textit{graph neural network}}
\acrodef{BiLSTM}{\textit{bidirectional long short-term memory}}

\newcommand{\OurModel}{CLIP-Siamese}

\author{Mariya Hendriksen}
 \affiliation{%
   \institution{AIRLab, University of Amsterdam}
     \city{}
  \country{}
 }
 \email{m.hendriksen@uva.nl}
 
 \author{Viggo Overes}
 \affiliation{%
   \institution{University of Amsterdam}
     \city{}
  \country{}
 }
 \email{mail@viggooveres.xyz}

\newcommand\sL{\ensuremath{\mathcal{L}}}

\newcommand\bh{\ensuremath{\mathbf{h}}}

\newcommand\bp{\ensuremath{\mathbf{p}}}

\newcommand\bx{\ensuremath{\mathbf{x}}}

\newcommand\BR{\ensuremath{\mathbb{R}}}

\usepackage{color}

\setcopyright{rightsretained}
\acmYear{2022}
\copyrightyear{2022}
\makeatletter
\renewcommand\@formatdoi[1]{\ignorespaces}
\makeatother
\acmISBN{}

\begin{document}

\title[Unimodal vs. Multimodal Siamese Networks for Outfit Completion]{Unimodal vs. Multimodal Siamese Networks\\ for Outfit Completion}

\renewcommand{\shortauthors}{Hendriksen and Overes}

\begin{abstract}
The popularity of online fashion shopping continues to grow. The ability to offer an effective recommendation to customers is becoming increasingly important. 
In this work, we focus on Fashion Outfits Challenge,  part of SIGIR 2022 Workshop on eCommerce.  The challenge is centered around \ac{FITB} task that implies predicting the missing outfit, given an incomplete outfit and a list of candidates.
In this paper, we focus on applying siamese networks on the task.
More specifically, we explore how combining information from multiple modalities (textual and visual modality) impacts the model's performance on the task.
We evaluate our model on the test split provided by the challenge organizers and the test split with gold assignments that we created during the development phase. We discover that using both visual, and visual and textual data demonstrates promising results on the task. We conclude by suggesting directions for further improvement of our method.
\end{abstract}



\maketitle


\section{Introduction}
\label{sec:introduction}

\begin{figure*}[!htb]
    \centering
    \includegraphics[width=0.7\textwidth]{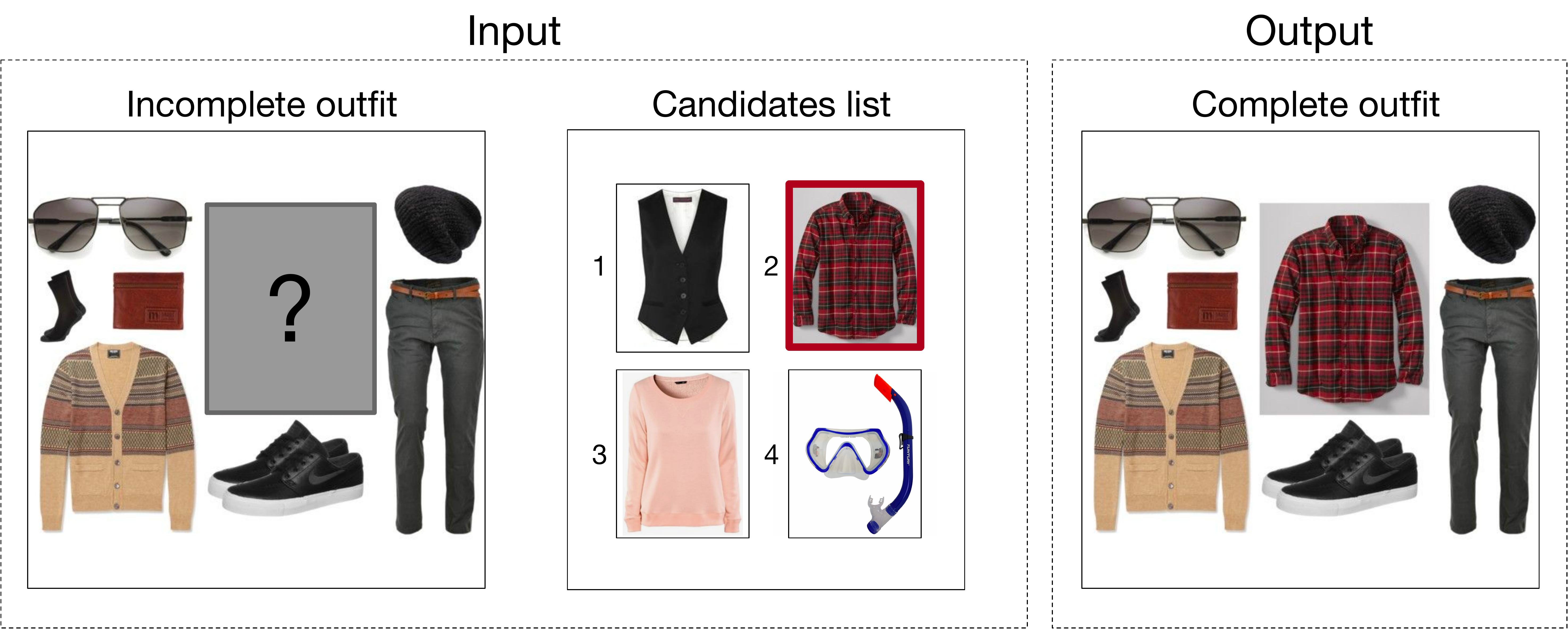}
    \caption{Example of \ac{FITB} task. Given an incomplete outfit and a list of candidates, we aim to select a candidate that would complete the outfit and return the complete outfit. In the example, the target item is the item \#2 in the candidates list.}
    \label{fig:task-example}
\end{figure*}

Fashion is becoming increasingly popular in modern e-commerce~\cite{fashion_ecom_2020}.
One of the common fashion recommendation tasks related to the problem is \ac{FITB}. The task consists of predicting a missing item, given an incomplete outfit, and a list of candidates. Figure~\ref{fig:task-example} illustrates the task.

\header{Fashion Outfits Challenge} In this work, we focus on \ac{FITB} task in the context of \textit{Fashion Outfits Challenge}~\footnote{\url{https://eval.ai/web/challenges/challenge-page/1721/overview}, Last accessed: 20.07.2022.}. The dataset consists of approximately \numprint{400000} products with product images and metadata. Besides, the dataset includes approximately \numprint{300000} outfits created by stylists and fashion experts. The models are evaluated via an online leaderboard on a test set. The gold standard assignments of the test set are not public. The metric used for performance evaluation is accuracy. Additionally, we evaluate the model performance on \ac{MRR}.

\header{Our solution} The main contributions of this work are as follows:
\begin{enumerate*}
    \item We apply the siamese network on the \ac{FITB} task and explore its effectiveness when using unimodal (Text or Image) and multimodal (Text \& Image) product representations.  We present a lightweight solution that uses only \numprint{697280} trainable parameters.
    \item We analyze the effectiveness and limitations of our method and discuss directions for future work.  We share our code and experimental settings to facilitate reproducibility of our results~\footnote{\url{https://github.com/mariyahendriksen/OutfitComposition}}.
\end{enumerate*}

\section{Related Work}
\label{sec:related_work}

\header{Outfit completion}
The majority of work on \ac{FITB} task was done on Polyvore dataset~\cite{polyvore_bilstm}. The authors of the dataset proposed to use \ac{BiLSTM} network on the task. The model leverages visual data alongside one-hot encoded product descriptions and treats the task as a sequence prediction problem.
\citet{cucurull_2019} propose to use \ac{GNN} on the \ac{FITB} task. In the work, they see each outfit as a graph and treat the outfit completion task as a missing link prediction problem.
\citet{Revanur_2021} propose to learn fashion compatibility in a semi-supervised way by learning pseudo positive and negative outfits while training the model. Another approach implies learning type-aware 
use type embeddings~
\citet{vasileva_2018} propose to jointly learn the notions of item similarity and compatibility while training the outfit completion model. \citet{veit_2015} propose to learn the compatibility of items using a siamese CNN trained on dyactic co-occurrences.
Unlike prior work in this domain, we investigate the performance of unimodal vs. multimodal siamese networks on the task of outfit completion.

\header{Multimodal fashion search} Multimodal fashion retrieval is an important and actively developing topic~\cite{hendriksen2022multimodal}.
Some of the related problems include fine-grained cross-modal retrieval~\cite{goei2021tackling}, machine translation~\cite{laenen2019multimodal}, and fashion recommendations~\cite{lin2019improving}

Unlike prior work in this domain, we focus on leveraging multimodal fashion product data on the \ac{FITB} task.


\section{Approach}
\label{sec:approach}

\header{Task definition} We follow the same notation as in~\citep{zhang2020contrastive,  hendriksen2022extending}. We present the input dataset as product-product pairs  $(\bx_p^i, \bx_p^j)$, where $\bx_p^i$ and $\bx_p^j$ represent information about two products.  A product-product pair $(\bx_p^i, \bx_p^j)$ is positive if both products belong to the same outfit; the pair is negative if products in the pair do not belong to the same outfit.
The product information includes images $\bx_i$, text $\bx_t$, and meta data,  i.e.,  $\bx_p = \{ \bx_i,  \bx_t, \bx_m \}$.

For the \ac{FITB} task,  we take as an input a list of products in an incomplete outfit and a list of candidate products; we aim to select a product that completes the outfit from the list of candidates.

\begin{figure*}[!htb]
    \centering
    \includegraphics[width=0.7\textwidth]{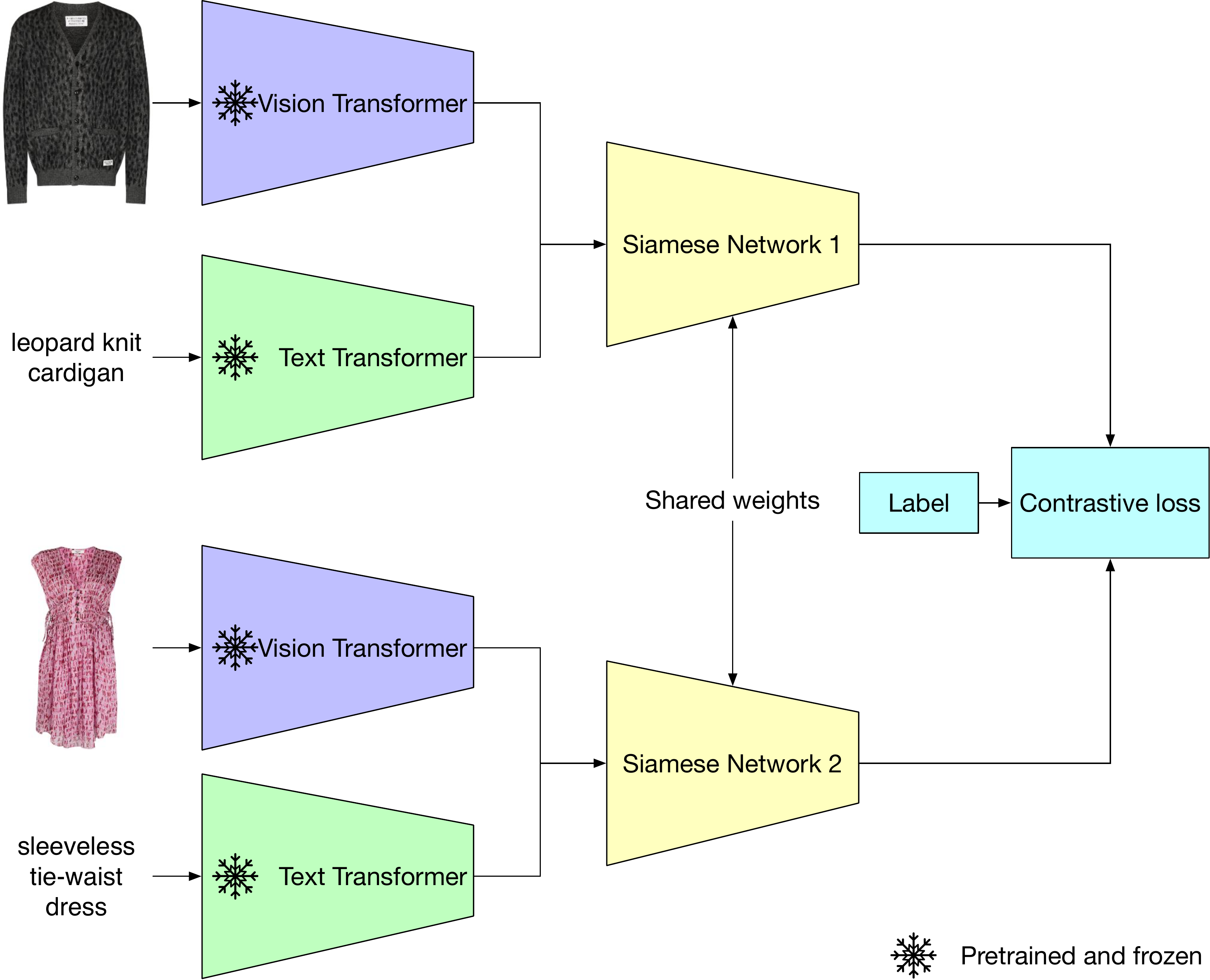}
    \caption{Overview of the proposed model.}
    \label{fig:multi-model}
\end{figure*}

\header{\OurModel{}} Figure~\ref{fig:multi-model} illustrates our approach. 
The model projects product information $\bx_p$ into a $d$-dimensional space with the resulting vector $\bp$.  The model consists of an encoding and a siamese modules.  It is trained with contrastive loss.

\header{Encoding product information}
We encode product textual and visual information with text and image encoder.
The \emph{image encoder} ($f_i$) takes as input a product image $\bx_i$.  The image $\bx_i$ is passed through the image encoder:
\begin{equation}
\bh_i = f_i (\bx_i).
\end{equation}

The \emph{text encoder} ($f_t$) takes a product textual information $\bx_t$ as input and returns a text representation $\bh_t$:
\begin{equation} 
\bh_t = f_t (\bx_t).
\end{equation}
To obtain the image and text representations, we use a pre-trained and frozen Vision Transformer and Text Transformer from CLIP model.

The image and text representations are passed to the \emph{siamese network} ($s_p$). The network takes as an input a concatenation of the image representation $\bh_i$ and text representation $\bh_t$,  and projects the resulting vector $\bh_p =  concat(\bh_i,  \bh_t) $ into the into a $d$-dimensional space:
\begin{equation}
\bp = g_p(\bh_p) = g_p(concat(\bh_i,  \bh_t))
\end{equation}
where $\bp \in \BR^d$.

\header{Loss function} After obtaining product representation for pair of products $(\bp_i, \bp_j)$, we use contrastive loss~\cite{hadsell2006dimensionality} to train \OurModel{}.  The loss goes over positive and negative product pairs.  Label $Y$ indicates if the pair is positive ($Y = 1$) or negative ($Y = 0$).
\begin{equation}
    \sL(Y, \bp_i, \bp_j) = (Y)D^2 + (1 - Y)\{\text{max}(0, m - D)\}^2
\end{equation}
where $D = D(\bp_i,  \bp_j) = || \bp_i - \bp_j ||_2$ is the euclidean distance,  $m > 0$ is a margin.


\section{Experimental Setup}
\label{sec:experiments}

\header{Metrics} We evaluate the model's performance using accuracy and \ac{MRR}.

\header{Baselines} We use category-based baseline provided by the challenge organizers, and CLIP~\cite{radford2021learning} as our baselines.

\header{Evaluation method} To explore how model performance changes w.r.t. unimodal vs. multimodal product representation, we train and evaluate \OurModel{} on three types of product representations:
\begin{enumerate*}
	\item \textit{Text}: we use only text data to build text-based product representations
	\item \textit{Image}: we use only product images to build image-based product representations
	\item \textit{Text \& Image}: we use both textual and visual product data to build multimodal product representations
\end{enumerate*} 

\header{Experiments} We run two experiments. 
In \emph{Experiment~1} we investigate how using unimodal and multimodal product representations impacts the accuracy of \OurModel{} when evaluated on the test split provided by the challenge organizers. 
We run the experiments on the test split provided by the challenge organizers,  and use accuracy as the metric. We use CLIP~\cite{radford2021learning} in zero-shot setting as our baseline.

In \emph{Experiment~2} we further investigate \OurModel{} performance with three different types of product representations.  We consider \ac{MRR} scores obtained by running the model on our own test split. Similar to the previous experiment, we use CLIP~\cite{radford2021learning} in a zero-shot setting as our baseline.


\section{Results}
\label{sec:results}

\header{Experiment 1: Fashion Outfits Challenge test split}
We investigate how using unimodal and multimodal product representations for training the model for the task impacts the accuracy of when we evaluate the model on the test split provided by the challenge organizers. We use CLIP in a zero-shot setting~\cite{radford2021learning} as a baseline. 

\begin{table}[H]
\centering
\begin{tabular}{lrrr}
\toprule
                                 & \multicolumn{3}{c}{\textbf{Accuracy}}                                                                       \\
                                 \cmidrule(r){2-4}
\multirow{-2}{*}{\textbf{Model}} & \textbf{Text}                     & \textbf{Image}                    & \multicolumn{1}{r}{\textbf{Text \& Image}} \\
\midrule
{CLIP zero-shot~\cite{radford2021learning}} & {\numprint{0.04160}} & {\numprint{0.04146}} & {\numprint{0.04246}} \\
{CLIP Siamese (Ours)} & {\numprint{0.04593}} & {\numprint{0.04864}} & {\textbf{\numprint{0.04920}}} \\
\bottomrule
\end{tabular}
\caption{Results of Experiment 1.  Models accuracy scores when using three different types of product representations. The best performance is highlighted in bold.}
\label{tab:fitb-acc}
\end{table}

The results are shown in Table~\ref{tab:fitb-acc}. In all cases, \OurModel{} outperforms CLIP zero-shot. The most significant relative gain is for image-based product representations where \OurModel{} outperforms CLIP zeros-shot by 17.32\%. It is followed by 15.86\% relative gain for text and image-based representations and 10.41\% gain for text-based representations.
Overall, \OurModel{} with text and image-based representations performs best.

\header{Experiment 2: Our own test split} To improve our understanding of model performance, we consider the performance in terms of \ac{MRR} scores. Since the the gold standard assignments for the test split is not released yet, we create our test split using scripts provided in \textit{utils} folder available on the challenge page.
\begin{table}[H]
\centering
\begin{tabular}{lrrr}
\toprule
                                 & \multicolumn{3}{c}{\textbf{\ac{MRR}}}                                                                       \\
                                 \cmidrule(r){2-4}
\multirow{-2}{*}{\textbf{Model}} & \textbf{Text}                     & \textbf{Image}                    & \multicolumn{1}{r}{\textbf{Text \& Image}} \\
\midrule
{CLIP zero-shot~\cite{radford2021learning}} & \numprint{0.16348} & \numprint{0.16140} & {\numprint{0,16277}} \\
{CLIP Siamese (Ours)} & {\numprint{0.16790}} & {\textbf{\numprint{0.18487}}} & {\numprint{0.18155}} \\
\bottomrule
\end{tabular}
\caption{Results of Experiment 2. Models \ac{MRR} scores when using three different types of product representations. The best performance is highlighted in bold.}
\label{tab:mrr}
\end{table}

Table~\ref{tab:mrr}, shows the experimental results for Experiment 2. Overall, \OurModel{} with image-based representations demonstrates the best performance, \OurModel{} with text and image-based representations is the second best.

\section{Conclusions}
\label{sec:conclusions}

In this paper, we present \OurModel{}, a model we created for Fashion Outfits Challenge. We evaluated the model on unimodal and multimodal product representations and showed that using both visual, and visual and textual data for building product representations demonstrates promising results. Future work includes further improvement of the model architecture and investigation of model performance on other datasets, e.g., Polyvore~\cite{polyvore_bilstm}.

\bibliographystyle{ACM-Reference-Format}
\bibliography{bibliography}
\balance

\end{document}